\documentclass[prd,twocolumn,groupedaddress,showpacs,nofootinbib]{revtex4}
\usepackage{graphicx}
\usepackage{dcolumn}
\usepackage{amssymb}
\usepackage{mathrsfs}
\usepackage{amsmath}
\usepackage{epsfig}
\usepackage[dvips]{color}
\usepackage{hhline}

\begin{document}

\title{Leptogenesis, radiative neutrino masses and inert Higgs triplet dark matter}

\author{Wen-Bin Lu}

\email{robinsonlu@sjtu.edu.cn}

\author{Pei-Hong Gu}

\email{peihong.gu@sjtu.edu.cn}

\affiliation{Department of Physics and Astronomy, Shanghai Jiao Tong University, 800 Dongchuan Road, Shanghai 200240, China}

\begin{abstract}

We extend the standard model by three types of inert fields including Majorana fermion singlets/triplets, real Higgs singlets/triplets and leptonic Higgs doublets. In the presence of a softly broken lepton number and an exactly conserved $Z_2^{}$ discrete symmetry, these inert fields together can mediate a one-loop diagram for a Majorana neutrino mass generation. The heavier inert fields can decay to realize a successful leptogenesis while the lightest inert field can provide a stable dark matter candidate. As an example, we demonstrate the leptogenesis by the inert Higgs doublet decays. We also perform a systematic study on the inert Higgs triplet dark matter scenario where the interference between the gauge and Higgs portal interactions can significantly affect the dark matter properties.

\end{abstract}


\maketitle

\section{Introduction}

The atmospheric, solar, accelerator and reactor neutrino experiments have established the phenomena of neutrino oscillations, which reveal that three flavors of neutrinos should be massive and mixed \cite{olive2014}. Hence we need new physics beyond the $SU(3)_c^{}\times SU(2)_L^{}\times U(1)^{}_{Y}$ standard model (SM) where the neutrinos are massless. Meanwhile, the cosmological observations indicate that the neutrino masses should be below the eV scale \cite{olive2014}. Along the lines to naturally generate the tiny neutrino masses, the famous seesaw \cite{minkowski1977} mechanism stands out as one of the most compelling paradigms. In the usual seesaw models \cite{minkowski1977,mw1980,flhj1989,ma1998,barr2003}, the neutrino masses are induced by some lepton-number-violating interactions, through which a lepton asymmetry could be produced and then converted to a baryon asymmetry by virtue of the sphaleron \cite{krs1985} processes. Thanks to this framework, one gets to understand the cosmic matter-antimatter asymmetry which is tantamount to the baryon asymmetry. This baryogensis scenario within the seesaw context is the well known leptogenesis \cite{fy1986} mechanism and has been extensively studied \cite{lpy1986,fps1995,ms1998,bcst1999,hambye2001,di2002,gnrrs2003,hs2004,bbp2005,dnn2008,dhh2014,ksy2015,fmmn2015}.

On the other hand, the existence of non-baryonic dark matter (DM) poses another challenge to particle physics and cosmology \cite{olive2014}. There have been a number of DM candidates in the literature. The particles for the DM may also play an essential role in the generation of the neutrino masses \cite{knt2003,ma2006,ma2006-2,kms2006,ms2009,chrt2011} and even the origin of the baryon asymmetry \cite{ma2006}. For example \cite{ma2006}, one extends the SM by two or more gauge-singlet fermions and a second iso-doublet Higgs scalar, which are odd under an exactly conserved $Z_2^{}$ discrete symmetry, to simultaneously explain the puzzles of the neutrino masses, the baryon asymmetry and the DM. Specifically, the new Higgs doublet can provide a real scalar to be a stable DM particle, while the new fermions with heavy Majorana masses can highly suppress the radiative neutrino masses and their decays can realize a successful leptogenesis.

In this paper, we will consider a class of models with Majorana fermion singlets/triplets, real Higgs singlets/triplets and leptonic Higgs doublets. Our models respect a softly broken lepton number and an exactly conserved $Z_2^{}$ discrete symmetry, so that they can only give the Majorana neutrino masses at one-loop level. The interactions for the neutrino mass generation can also allow the decays of the heavier non-SM fields to produce a lepton asymmetry stored in the SM leptons and then realize the leptogenesis. Furthermore, the lightest non-SM field can provide a stable DM candidate. As an example, we will demonstrate the leptogenesis scenario by the inert Higgs doublet decays. We will also perform a systematic study on the inert Higgs triplet DM scenario where the Higgs portal interaction can significantly affect the dark matter annihilation and scattering.

\section{The models}

The non-SM gauge-singlet/iso-triplet fermions, gauge-singlet/iso-triplet Higgs scalars and iso-doublet Higgs scalars are denoted by
\begin{eqnarray}
\label{nsmf}
&&N_{R}^{}(1,1,0)/T_{L}^{}(1,3,0)=\left[\begin{array}{cc} T^0_{L}/\sqrt{2} & T^+_{L} \\
[2mm] T^-_{L} & -T^0_{L}/\sqrt{2} \end{array}\right]\,;\nonumber\\
[2mm]
&&\chi(1,1,0)/\Sigma(1,3,0)=\left[\begin{array}{cc} \Sigma^0_{}/\sqrt{2} & \Sigma^+_{} \\
[2mm] \Sigma^-_{} & -\Sigma^0_{}/\sqrt{2} \end{array}\right]=\Sigma^\dagger_{}\,;\nonumber\\
[2mm]&&
\eta(\!\!\begin{array}{c}1,2,-\frac{1}{2}\end{array}\!\!)=\left[\begin{array}{c} \eta^0_{} \\
[2mm] \eta^-_{} \end{array}\right]=\left[\begin{array}{c} \frac{1}{\sqrt{2}}(\eta^0_{R}+i\eta^0_{I}) \\
[2mm] \eta^-_{} \end{array}\right]\,.
\end{eqnarray}
Here and thereafter the brackets following the fields describe the transformations under the $SU(3)_c^{}\times SU(2)_L^{}\times U(1)^{}_{Y}$ gauge groups. Our models also contain a $Z_2^{}$ discrete symmetry under which these non-SM fields are odd while the SM fields are even, i.e.
\begin{eqnarray}
\!\!\!\!\!\!\!\!&&(N_R^{}/T_L^{};\chi/\Sigma;\eta) \stackrel{Z_2^{}}{\longrightarrow}-(N_R^{}/T_L^{};\chi/\Sigma;\eta)\,,~~\textrm{SM}\stackrel{Z_2^{}}{\longrightarrow}\textrm{SM}\,.\nonumber\\
\!\!\!\!\!\!\!\!&&
\end{eqnarray}
Among the non-SM fields, the Higgs doublets $\eta$ carry a lepton number the same as that of the SM leptons, while the others do not.

We require the $Z_2^{}$ symmetry to be exactly conserved while allowing the lepton number to be softly broken.  With all the considerations stated, four options to extend the SM emerge:
\begin{itemize}
\item the fermion singlets + Higgs singlets + Higgs doublets (SSD) model,
\begin{eqnarray}
\label{ssd}
\mathcal{L}_{\textrm{SSD}}^{}&\supset&-\frac{1}{2}(M_{N}^{})_{ij}^{}\bar{N}_{Ri}^{}N_{Rj}^{c}
-(M_{\eta}^2)_{ij}^{}\eta^\dagger_i\eta^{}_j\nonumber\\
&&-\rho_{ij}^{}\eta^\dagger_{i}\chi_j^{}\phi-y_{\alpha ij}^{}\bar{l}_{L\alpha}^{}N_{Ri}^{}\eta_j^{}+\textrm{H.c.}\nonumber\\
&&-\frac{1}{2}(M^2_\chi)_{ij}^{}\chi_i^{}\chi_j^{}\,.
\end{eqnarray}
\item the fermion singlets + Higgs triplets + Higgs doublets (STD) model,
\begin{eqnarray}
\label{std}
\mathcal{L}_{\textrm{STD}}^{}&\supset&-\frac{1}{2}(M_{N}^{})_{ij}^{}\bar{N}_{Ri}^{}N_{Rj}^{c}
-(M_{\eta}^2)_{ij}^{}\eta^\dagger_i\eta^{}_j\nonumber\\
&&-\sqrt{2}\rho_{ij}^{}\eta^\dagger_{i}\Sigma_j^{}\phi
-y_{\alpha ij}^{}\bar{l}_{L\alpha}^{}N_{Ri}^{}\eta_j^{}+\textrm{H.c.}\nonumber\\
&&-\frac{1}{2}(M^2_\Sigma)_{ij}^{}\textrm{Tr}(\Sigma_i^{}\Sigma_j^{})\,.
\end{eqnarray}
\item the fermion triplets + Higgs singlets + Higgs doublets (TSD) model,
\begin{eqnarray}
\label{tsd}
\mathcal{L}_{\textrm{TSD}}^{}&\supset&-\frac{1}{2}(M_{T}^{})_{ij}^{}\textrm{Tr}(\bar{T}_{Li}^{c}i\tau_2^{}T_{Lj}^{}i\tau_2^{})
-(M_{\eta}^2)_{ij}^{}\eta^\dagger_i\eta^{}_j\nonumber\\
&&-\rho_{ij}^{}\eta^\dagger_{i}\phi\chi_j^{}
-\sqrt{2}y_{\alpha ij}^{}\bar{l}_{L\alpha}^{}i\tau_2^{}T_{Li}^{c}i\tau_2^{}\eta_j^{}+\textrm{H.c.}\nonumber\\
&&-\frac{1}{2}(M^2_\chi)_{ij}^{}\chi_i^{}\chi_j^{}\,.
\end{eqnarray}
\item the fermion triplets + Higgs triplets + Higgs doublets (TTD) model,
\begin{eqnarray}
\label{ttd}
\mathcal{L}_{\textrm{TTD}}^{}&\supset&-\frac{1}{2}(M_{T}^{})_{ij}^{}\textrm{Tr}(\bar{T}_{Li}^{c}i\tau_2^{}T_{Lj}^{}i\tau_2^{})
-(M_{\eta}^2)_{ij}^{}\eta^\dagger_i\eta^{}_j\nonumber\\
&&-\sqrt{2}\rho_{ij}^{}\eta^\dagger_{i}\Sigma_{j}^{} \phi
-\sqrt{2}y_{\alpha ij}^{}\bar{l}_{L\alpha}^{}i\tau_2^{}T_{Li}^{c}i\tau_2^{}\eta_j^{}+\textrm{H.c.}\nonumber\\
&&-\frac{1}{2}(M^2_\Sigma)_{ij}^{}\textrm{Tr}(\Sigma_i^{}\Sigma_j^{})\,.
\end{eqnarray}
\end{itemize}
Here $\phi$ and $l_{L}^{}$ are the SM Higgs and lepton doublets,
\begin{eqnarray}
\phi(1,2,-\frac{1}{2})&=&\left[\begin{array}{c} \phi^{0}_{} \\
[2mm] \phi^{-}_{}\end{array}\right]\,,\nonumber\\
l_{L\alpha}^{}(1,2,-\frac{1}{2})&=&\left[\begin{array}{c} \nu^{}_{L\alpha} \\
[2mm] e_{L\alpha}^{}\end{array}\right]~~(\alpha=e\,,~\mu\,,~\tau)\,.
\end{eqnarray}
For simplicity, we shall not write down the full SM Lagrangian where the Higgs doublet $\phi$ has the potential as below,
\begin{eqnarray}
\label{sm}
\mathcal{L}_{\textrm{SM}}^{}\supset -\mu_\phi^2\phi^\dagger_{}\phi -\lambda(\phi^\dagger_{}\phi)^2_{}\,.
\end{eqnarray}

We would like to emphasize that the lepton number is only allowed to be softly broken by the cubic coupling among the SM Higgs scalar $\phi$ and the non-SM Higgs scalars $\eta$ and $\chi/\Sigma$, i.e. the $\rho$-term in Eqs. (\ref{ssd}-\ref{ttd}). Meanwhile, the exactly conserved $Z_2^{}$ discrete symmetry will not be broken at any scales and hence the non-SM Higgs scalars will not develop any nonzero vacuum expectation values (VEVs). This $Z_2^{}$ symmetry has also forbidden the other gauge invariant terms involving the non-SM fields. In this sense, we will refer to the non-SM fields (\ref{nsmf}) as the inert fermion singlets/triplets, the inert Higgs singlets/triplets and the inert Higgs doublets, respectively.

Without loss of generality and for the sake of convenience, we can choose the basis in which the Majorana mass matrix of the inert fermion singlets $N_R^{}$  is real and diagonal, i.e.
\begin{eqnarray}
M_N^{}=\textrm{diag}\{M_{N_1}^{},...\}~~\textrm{with}~~M_{N_1}^{}<M_{N_2}^{}<...\,.
\end{eqnarray}
Accordingly we can define the Majorana fermions as below,
\begin{eqnarray}
N_i^{}=N_{R_i}^{}+N_{R_i}^c=N_i^c\,.
\end{eqnarray}
Similar procedures for the inert fermion triplets $T_L^{}$ lead to
\begin{eqnarray}
M_T^{}=\textrm{diag}\{M_{T_1}^{},...\}~~\textrm{with}~~M_{T_1}^{}<M_{T_2}^{}<...\,,
\end{eqnarray}
and hence the physical states,
\begin{eqnarray}
T_i^{0}&=&T_{L_i}^{0}+(T_{L_i}^{0})^c_{}=(T_i^0)^c_{}\,,\nonumber\\
T_i^{\pm}&=&T_{L_i}^{\pm}+(T_{L_i}^{\mp})^c_{}=(T_i^\mp)^c_{}\,.
\end{eqnarray}
For the same reason, we can rotate the inert Higgs scalars $\chi/\Sigma$ and $\eta$ to diagonalize their mass terms,
\begin{eqnarray}
M_\chi^2&=&\textrm{diag}\{M_{\chi_1}^{2},...\}~~\textrm{with}~~M_{\chi_1}^{2}<M_{\chi_2}^{2}<...\,,\nonumber\\
M_\Sigma^2&=&\textrm{diag}\{M_{\Sigma_1}^{2},...\}~~\textrm{with}~~M_{\Sigma_1}^{2}<M_{\Sigma_2}^{2}<...\,;\nonumber\\
M_\eta^2&=&\textrm{diag}\{M_{\eta_1}^{2},...\}~~\textrm{with}~~M_{\eta_1}^{2}<M_{\eta_2}^{2}<...\,.
\end{eqnarray}

\section{Inert Higgs triplet dark matter}

The present models will invariantly select a stable particle from the inert fields. Since this stable particle leaves a relic density in the universe, it should be neutral and could be a viable DM candidate.

Firstly, we consider the fermionic DM. In the SSD and STD models, the lightest inert fermion singlet $N_{R}^{}$ can be the DM particle \cite{knt2003,ma2006,ma2006-2,kms2006}. In this case, the DM fermion can annihilate into the SM leptons through the t-channel exchange of the inert Higgs doublets. Therefore, the inert Higgs doublets cannot be too heavy while the Yukawa couplings cannot be too small. The detailed studies can be found in \cite{bglz2009}. In the TSD and TTD models, the neutral component $T_{L}^0$ of the lightest inert fermion triplet $T_{L}^{}$ can be the DM particle \cite{ms2009,cfs2006}. Since the DM fermion now have the $SU(2)_L^{}$ gauge couplings, its annihilation can be free of the Yukawa interactions. This means the inert Higgs doublets can be very heavy and the Yukawa couplings can be quite small, which indeed falls in the context of the minimal DM scenario \cite{cfs2006}.

Alternatively, the DM particle can be a scalar from the inert Higgs singlets/triplets and doublets. In fact, after the SM Higgs scalar $\phi$ develops its VEV to spontaneously break the electroweak symmetry, it can be written as
\begin{eqnarray}
\phi=\left[\begin{array}{c}
\frac{1}{\sqrt{2}}(h+v )\\
[2mm]0\end{array}\right]\,,
\end{eqnarray}
with $h$ being the Higgs boson and $v\simeq 246\,\textrm{GeV}$ being the VEV. This gives rise to the mixing between the inert Higgs scalars $\chi/\Sigma$ and $\eta$ due to the $\rho$-term in Eqs. (\ref{ssd}-\ref{ttd}), from which the DM scalar should be the lightest mass eigenstate obtained. To be simple and instructive, we hereby will only consider some limiting cases where the mixings could be essentially ignored.

If the inert Higgs singlets $\chi$ or triplets $\Sigma$ are much heavier than the lightest inert Higgs doublet $\eta$, we actually arrive at the inert Higgs doublet DM scenario which has been studied in a lot of literature \cite{ma2006,cfs2006,bhr2006,hllr2009}. In the usual inert Higgs doublet DM scenario, the inert Higgs doublet $\eta$ has a quartic coupling with the SM Higgs doublet $\phi$, i.e.
\begin{eqnarray}
\mathcal{L}\supset-\epsilon(\eta^\dagger_{}\phi)^2_{}+\textrm{H.c.}\,.
\end{eqnarray}
This term will induce the required mass split between the real and imaginary parts of $\eta$'s neutral component after the electroweak symmetry breaking. By integrating out the heavy inert Higgs singlets $\chi$ in the SSD and TSD models, or the heavy inert Higgs triplets $\Sigma$ in the STD and TTD models, we can obtain the coupling,
\begin{eqnarray}
\epsilon=-\rho\frac{1}{M_{\chi,\Sigma}^2}\rho^T_{}\,.
\end{eqnarray}

In the SSD and TSD models, the lightest inert Higgs singlet $\chi$ can dominate the DM scalar if it is much lighter than the inert Higgs doublets $\eta$. Consequently, the gauge interactions of the DM scalar are negligible compared to the more significant Higgs portal interaction between the DM scalar and the SM Higgs scalar. This simple DM scenario has attracted many people \cite{sz1985,mcdonald1994}.

In the STD and TTD models, the lightest inert Higgs triplet $\Sigma$ can be set much lighter than the inert Higgs doublets $\eta$. In this case we can work with the inert Higgs triplet $\Sigma$ as an approximately physical state, i.e.
\begin{eqnarray}
\hat{\Sigma}^0_{}\simeq \Sigma^0_{}\,,~~\hat{\Sigma}^\pm_{}\simeq \Sigma^\pm_{}\,.
\end{eqnarray}
The $SU(2)_L^{}$ radiative corrections will induce a mass split, making the charged components slightly heavier than the neutral one \cite{cfs2006},
\begin{eqnarray}
m_{\Sigma}^{}\gg \Delta m &=& m_{\Sigma^{\pm}_{}}-m_{\Sigma^0_{}}^{}\
\nonumber\\
&=&\frac{g^2_{}m_\Sigma^{}}{16\pi^2_{}}\left[f\left(\frac{m_W^{}}{m_\Sigma}\right)
-\cos^2_{}\theta_W^{} f\left(\frac{m_Z^{}}{m_\Sigma}\right)\right]\,,\nonumber\\
&&
\end{eqnarray}
where
\begin{eqnarray}
\!\!\!f(r)\!\!\!&=&\!\!\!\left\{\!\!\!\begin{array}{lll}-\frac{1}{2}\!\!\left[r^4_{}\ln r-r(r^2-4)^{\frac{3}{2}}_{}\ln \frac{r+\sqrt{r^2-4}}{2}\right]\!\!&
\textrm{for}\!\!&r\geq 2\,,\\
[2mm]
+\frac{1}{2}\!\!\left[r^4_{}\ln r+r(4-r^2)^{\frac{3}{2}}_{}\arctan\frac{\sqrt{4-r^2}}{r}\right]\!\!&
\textrm{for}\!\!&r\leq 2\,.\end{array}\right.\nonumber\\
\!\!\!\!\!\!&&\!\!\!
\end{eqnarray}
For $m_\Sigma^{}\gg m_{Z,W}^{}$, the radiative mass split $\Delta m$ can arrive at a determined value,
\begin{eqnarray}
\label{msplit}
\Delta m=\frac{g^2_{}}{4\pi}\sin^2_{}\left(\frac{\theta_W^{}}{2}\right)m_W^{}=167\,\textrm{MeV}\,.
\end{eqnarray}
The $\Sigma^\pm_{}-\Sigma^0_{}$ mass difference can be also induced at tree level due to the mixing between the inert Higgs triplet $\Sigma$ and the inert Higgs doublets $\eta$. However, we have checked that this tree-level contribution should be far below the radiative corrections provided that the inert Higgs doublets $\eta$ are much heavier than the inert Higgs triplet $\Sigma$. With the radiative mass split (\ref{msplit}), the charged $\Sigma^{\pm}_{}$ can decay into the neutral $\Sigma^0_{}$ with a virtual $W^{\pm}_{}$ before the Big Bang Nucleosynthesis (BBN) epoch. Therefore, the stable $\Sigma^{0}_{}$ can serve as the DM particle.

In the following we shall perform a systematic study on this inert Higgs triplet DM scenario \cite{cfs2006,fprw2009,agn2011}. Particular emphasis will be placed on investigating some interesting implications arising from the Higgs portal interaction between the inert Higgs triplet $\Sigma$ and the SM Higgs doublet $\phi$, i.e.
\begin{eqnarray}
\mathcal{L}_{\textrm{STD}/\textrm{TTD}}^{}\supset-\frac{1}{2}\kappa_1^{}\phi^\dagger_{}\phi\textrm{Tr}\Sigma_{}^{2}
-\frac{1}{2}\kappa_3^{}\phi^\dagger_{}\Sigma^2_{}\phi\,.
\end{eqnarray}
Subjected to the stability and perturbativity requirements, the Higgs portal coupling $\kappa$ should be in the range as below,
\begin{eqnarray}
&&-2.6\lesssim  -2\sqrt{\lambda\xi}\leq \kappa\equiv \kappa_1^{}+\frac{1}{2}\kappa_3^{}<4\pi~~\textrm{for}\nonumber\\
&&\xi\equiv \xi_1^{}+\frac{1}{2}\xi_3^{}<4\pi\,,~~\lambda\simeq 0.13\,.
\end{eqnarray}
Here the new parameters $\xi_{1,3}^{}$ are the self quartic couplings of the Higgs triplet $\Sigma$, i.e.
\begin{eqnarray}
\mathcal{L}_{\textrm{STD}/\textrm{TTD}}^{}\supset-\frac{1}{4}\xi_1^{}[\textrm{Tr}(\Sigma_{}^{2})]^{2}_{}
-\frac{1}{4}\xi_3^{}\textrm{Tr}(\Sigma^4_{})\,.
\end{eqnarray}
As for the choice $\lambda\simeq 0.13$, it is given by
\begin{eqnarray}
\label{lambda}
\lambda\simeq \frac{m_h^2}{2v^2_{}}\simeq 0.13 ~~\textrm{for}~~m_h^{}=125\,\textrm{GeV}\,,~v=246\,\textrm{GeV}\,.
\end{eqnarray}
We will explain this point later.

\subsection{Dark matter mass}

Since the $\Sigma^{\pm}_{}$ and $\Sigma^0_{}$ scalars now are highly quasi-degenerate, all the following annihilation and co-annihilation channels,
\begin{eqnarray}
\Sigma^{0}_{}\Sigma^{0}_{}&\rightarrow& W^{+}_{}W^{-}_{}\,,~~\phi^{0\ast}_{}\phi^{0}_{}\,,~~\phi^{+}_{}\phi^{-}_{}\,;\nonumber\\
\Sigma^{0}_{}\Sigma^{\pm}_{}&\rightarrow& W^{3}_{}W^{\pm}_{}\,,~~f'\bar{f}\,,~~\phi^0_{}\phi^{+}_{}\,,~~\phi^{0\ast}_{}\phi^{-}_{}\,;\nonumber\\
\Sigma^{\pm}_{}\Sigma^{\mp}_{}&\rightarrow& W^{\pm}_{}W^{\mp}_{}\,,~~f\bar{f}\,,~~\phi^{0\ast}_{}\phi^{0}_{}\,,~~\phi^{+}_{}\phi^{-}_{}\,;\nonumber\\
\Sigma^{\pm}_{}\Sigma^{\pm}_{}&\rightarrow& W^{\pm}_{}W^{\pm}_{}\,,
\end{eqnarray}
should be taken into account in determining the relic density of the DM particle $\Sigma^0_{}$ \cite{gs1991,gg1991}. Here $f$ and $f'$ denote the SM fermions. Up to the $p$-wave contributions, we calculate the thermally averaged cross sections of the above annihilations and co-annihilations,
\begin{eqnarray}
\langle\sigma_{\Sigma^{0}_{}\Sigma^{0}_{}}^{} v_{\textrm{rel}}^{}\rangle&=&\frac{g_{}^4}{4\pi m_\Sigma^2}\left(1-\frac{5}{x}\right)+\frac{\kappa^2_{}}{16\pi m_\Sigma^2}\left(1-\frac{3}{x}\right)\,, \nonumber \\
\langle\sigma_{\Sigma^{0}_{}\Sigma^{\pm}_{}}^{} v_{\textrm{rel}}^{}\rangle&=&\frac{g_{}^4}{16\pi m_\Sigma^2}\left(1-\frac{5}{4x}\right) \,,\nonumber \\
\langle\sigma_{\Sigma^{+}_{}\Sigma^{-}_{}}^{} v_{\textrm{rel}}^{}\rangle&=&\frac{3g_{}^4}{16\pi m_\Sigma^2}\left(1-\frac{51}{4x}\right)+\frac{\kappa^2_{}}{16\pi m_\Sigma^2} \left(1-\frac{3}{x}\right)\,, \nonumber \\
\langle\sigma_{\Sigma^{\pm}_{}\Sigma^{\pm}_{}}^{} v_{\textrm{rel}}^{}\rangle&=&\frac{g_{}^4}{8\pi m_\Sigma^2}\left(1-\frac{5}{x}\right)\,,
\end{eqnarray}
and then obtain an effective cross section \cite{st2003},
\begin{eqnarray}
\label{effsection}
\langle\sigma_{\textrm{eff}}^{} v_{\textrm{rel}}^{}\rangle&=&\frac{1}{9}\langle\sigma_{\Sigma^{0}_{}\Sigma^{0}_{}}^{} v_{\textrm{rel}}^{}\rangle+
\frac{4}{9}\langle\sigma_{\Sigma^{0}_{}\Sigma^{\pm}_{}}^{} v_{\textrm{rel}}^{}\rangle\nonumber\\
&&+
\frac{2}{9}\langle\sigma_{\Sigma^{+}_{}\Sigma^{-}_{}}^{} v_{\textrm{rel}}^{}\rangle+
\frac{2}{9}\langle\sigma_{\Sigma^{\pm}_{}\Sigma^{\pm}_{}}^{} v_{\textrm{rel}}^{}\rangle\nonumber\\
&=&\frac{g_{}^4}{8\pi m_\Sigma^2}\left(1-\frac{47}{12 x}\right)+\frac{\kappa^2_{}}{48\pi m_\Sigma^2}\left(1-\frac{3}{x}\right)\,,\nonumber\\
&&
\end{eqnarray}
where we have defined
\begin{eqnarray}
x\equiv \frac{m_\Sigma^{}}{T}\,.
\end{eqnarray}
The DM relic density then can be well approximated to \cite{gs1991,kt1990}
\begin{eqnarray}\label{relicdensity}
\Omega_{\textrm{DM}}^{}h^2_{}\simeq\frac{1.07\times10^9_{}\textrm{GeV}^{-1}_{}}{J(x_F^{}) g_{\ast}^{1/2} M_{\textrm{Pl}}^{}}\,,
\end{eqnarray}
with $M_\textrm{Pl}^{}\simeq 1.22\times 10^{19}\,\textrm{GeV}$ being the Planck mass, $g_{\ast}^{}\simeq 106.75$ being the number of the relativistic degrees of freedom at the freeze-out point, while $J(x_F^{})$ being an integral,
\begin{eqnarray}
J(x_F^{})&=&\int_{x_F^{}}^\infty \frac{\langle\sigma_{\textrm{eff}}^{}v_{\textrm{rel}}^{}\rangle}{x^2_{}} dx\,,
\end{eqnarray}
determined by the freeze-out point,
\begin{eqnarray}
\label{freezeout}
x_F^{}=\ln \frac{3\times 0.038\,M_{\textrm{Pl}}^{}m_\Sigma^{}\langle\sigma_{\textrm{eff}}^{}v_{\textrm{rel}}^{}\rangle}{g_\ast^{1/2} x_F^{1/2}}\,,
\end{eqnarray}
at which the annihilations and co-annihilations become slower than the expansion rate of the universe.

\begin{figure}[htp]
  \vspace{0cm} \centering
  \includegraphics[width=8cm]{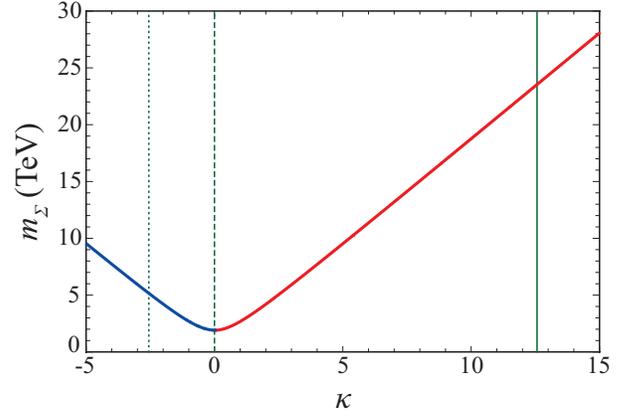}
  \caption{The correlation between the dark matter mass $m_\Sigma^{}$ and the Higgs portal coupling $\kappa$. The dot, dash and solid vertical lines correspond to $\kappa=-2.6$, $0$ and $4\pi$, respectively. We have $m_\Sigma^{}=5.2\,\textrm{TeV}$, $m_\Sigma^{}=2\,\textrm{TeV}$ and $m_\Sigma^{}=23.6\,\textrm{TeV}$ for $\kappa=-2.6$, $\kappa=0$ and $\kappa=4\pi$, respectively.}
  \label{mass-coupling}
\end{figure}

From Eqs. (\ref{effsection}-\ref{freezeout}), we can easily understand that the present DM relic density $\Omega_{\text{DM}}h^2=0.1188\pm0.0010$ \cite{ade2015} only depends on two parameters: the mass of the DM scalar $\Sigma^0_{}$ and the Higgs portal coupling between the inert Higgs triplet $\Sigma$ and the SM Higgs doublet $\phi$. In Fig. \ref{mass-coupling}, we show the correlation between the DM mass $m_\Sigma^{}$ and the Higgs portal coupling $\kappa$. Specifically, $m_\Sigma^{}$ will decrease from $5.2\,\textrm{TeV}$ to $2\,\textrm{TeV}$ when $\kappa$ increases from $-2.6$ to $0$, subsequently, $m_\Sigma^{}$ will increase to $23.6\,\textrm{TeV}$ when $\kappa$ increases to $4\pi$.

\subsection{Dark matter direct detection}

\begin{figure*}
\vspace{4cm} \epsfig{file=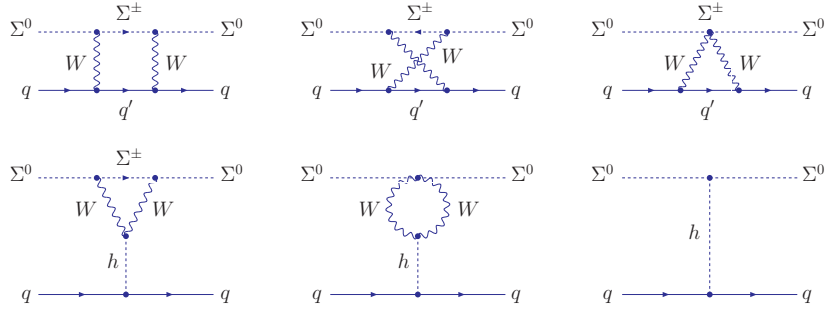, bbllx=5.7cm, bblly=6.0cm,
bburx=15.7cm, bbury=16cm, width=5.5cm, height=5.5cm, angle=0,
clip=0} \vspace{-5cm} \caption{\label{dm-nucleon} The effective couplings of the dark matter scalar to the SM quarks. The tree-level effect is only induced by the Higgs portal interaction.}
\end{figure*}

\begin{figure*}[htp]
  \vspace{0.75cm} \centering
  \includegraphics[width=15cm]{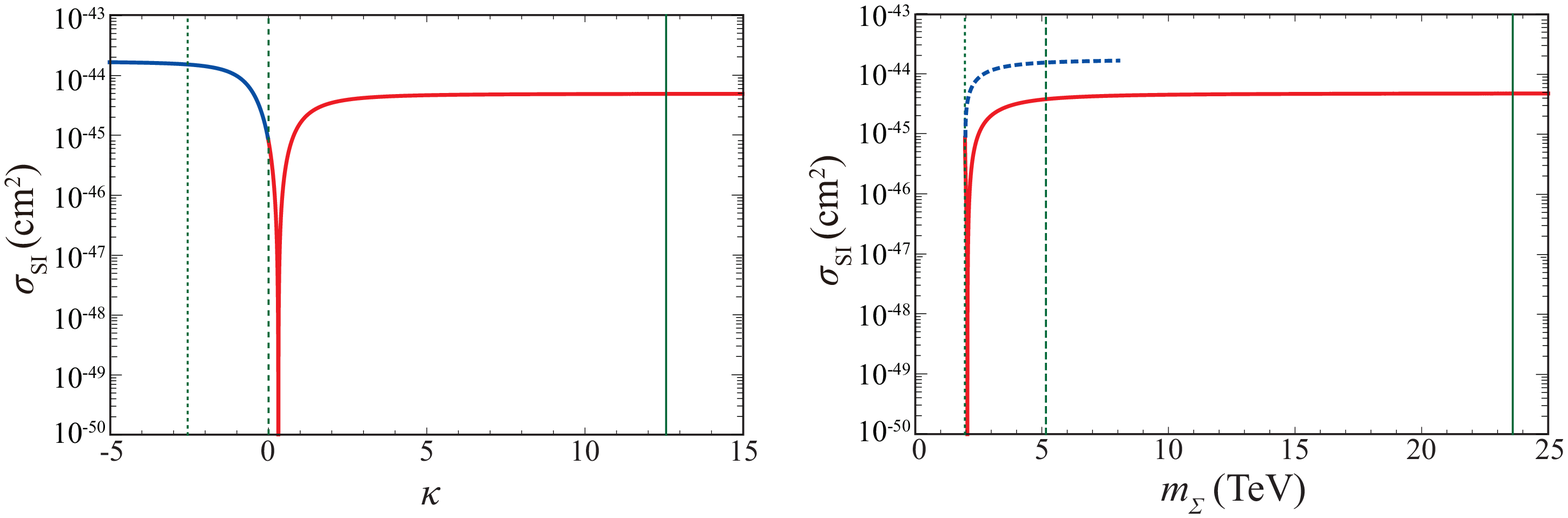}
  \caption{The dependence of the DM-nucleon scattering cross section $\sigma_{\textrm{SI}}^{}$ on the Higgs portal coupling $\kappa$ and the DM mass $m_\Sigma^{}$. In the left panel, the dot, dash and solid vertical lines correspond to $\kappa=-2.6$, $\kappa=0$ and $\kappa=4\pi$, respectively. In the right panel, the dot curve is for $\kappa<0$ and the solid curve is for $\kappa>0$, while the dot, dash and solid vertical lines correspond to $m_\Sigma^{}=2\,\textrm{TeV}$, $m_\Sigma^{}=5.2\,\textrm{TeV}$ and $m_\Sigma^{}=23.6\,\textrm{TeV}$, respectively. We have $\sigma_{\textrm{SI}}^{}=1.8\times 10^{-44}_{}\,\textrm{cm}^2_{}$ for $\kappa=-2.6$ and $m_\Sigma^{}=5.2\,\textrm{TeV}$, $\sigma_{\textrm{SI}}^{}=0.9\times 10^{-45}_{}\,\textrm{cm}^2_{}$ for $\kappa=0$ and $m_\Sigma^{}=2\,\textrm{TeV}$, while $\sigma_{\textrm{SI}}^{}=5\times 10^{-45}_{}\,\textrm{cm}^2_{}$ for $\kappa=4\pi$ and $m_\Sigma^{}=23.6\,\textrm{TeV}$. }
  \label{scattering-section}
\end{figure*}

As shown in Fig. \ref{dm-nucleon}, the DM scalar $\Sigma^0_{}$ can scatter off a nuclei at tree and loop level. Note the tree-level effect is only induced by the Higgs portal interaction between the inert Higgs triplet $\Sigma$ and the SM Higgs doublet $\phi$. We have performed an improved computation on the spin-independent DM-nucleon scattering cross section incorporating all interfering channels,
\begin{eqnarray}
\label{si}
\sigma_{\textrm{SI}}^{}&=&\frac{g^8_{}}{256\pi^3_{}} \frac{f_N^2 m_N^4}{m_W^2} \left[\frac{1}{m_W^2}+\frac{1}{m_h^2} \left( 1-\frac{16\pi\kappa}{g^4_{}}\frac{m_W^{}}{m_\Sigma^{}}\right)
\right]^2\nonumber\\
&&\textrm{for}~~m_\Sigma^{} \gg m_W^{} \gg m_N^{}\,.
\end{eqnarray}
Here $m_N^{}\simeq 1\,\textrm{GeV}$ is the nucleon mass, $f_N^{}\simeq 0.3$ \cite{efo2000} is the effective coupling of the Higgs boson to the nucleon. The cross section $\sigma_{\textrm{SI}}^{}$ is a function of the two correlated parameters: the DM mass $m_\Sigma^{}$ and the Higgs portal coupling $\kappa$. Remarkably, Eq. (\ref{si}) gives a zero point,
\begin{eqnarray}
\sigma_{\textrm{SI}}^{}= 0 \Rightarrow \kappa= \frac{g^4_{}m_\Sigma^{}}{16\pi m_W^{}}\left(1+\frac{m_h^2}{m_W^2}\right)\,.
\end{eqnarray}
Although the above extreme condition would not be exactly accessible constrained by the correlation between the DM mass $m_\Sigma^{}$ and the coupling $\kappa$, our result indeed exhibits the intriguing property that the spin-independent cross section $\sigma_{\textrm{SI}}^{}$ might be highly suppressed for some choice of $\kappa$ and $m_\Sigma^{}$.

In Fig. \ref{scattering-section}, we show the dependence of the DM-nucleon scattering cross section $\sigma_{\textrm{SI}}^{}$ on the Higgs portal coupling $\kappa$ and the DM mass $m_\Sigma^{}$. We find the coupling $\kappa$ can significantly affect the cross section $\sigma_{\textrm{SI}}^{}$. For example, we read $\sigma_{\textrm{SI}}^{}\simeq 1.8\times 10^{-44}_{}\,\textrm{cm}^2_{}$, $0.9\times 10^{-45}\,\textrm{cm}^2_{}$ and $5\times 10^{-45}_{}\,\textrm{cm}^2_{}$ for $\kappa=-2.6$, $0$ and $4\pi$, respectively. The cross section $\sigma_{\textrm{SI}}^{}$ could even drastically decrease to an extremely small value for $\kappa \rightarrow 0.3$ or $m_\Sigma^{}\rightarrow 2.05\,\textrm{TeV}$.

\subsection{Higgs phenomenology}

\begin{figure*}[htp]
 \vspace{0cm} \centering
  \includegraphics[width=15cm]{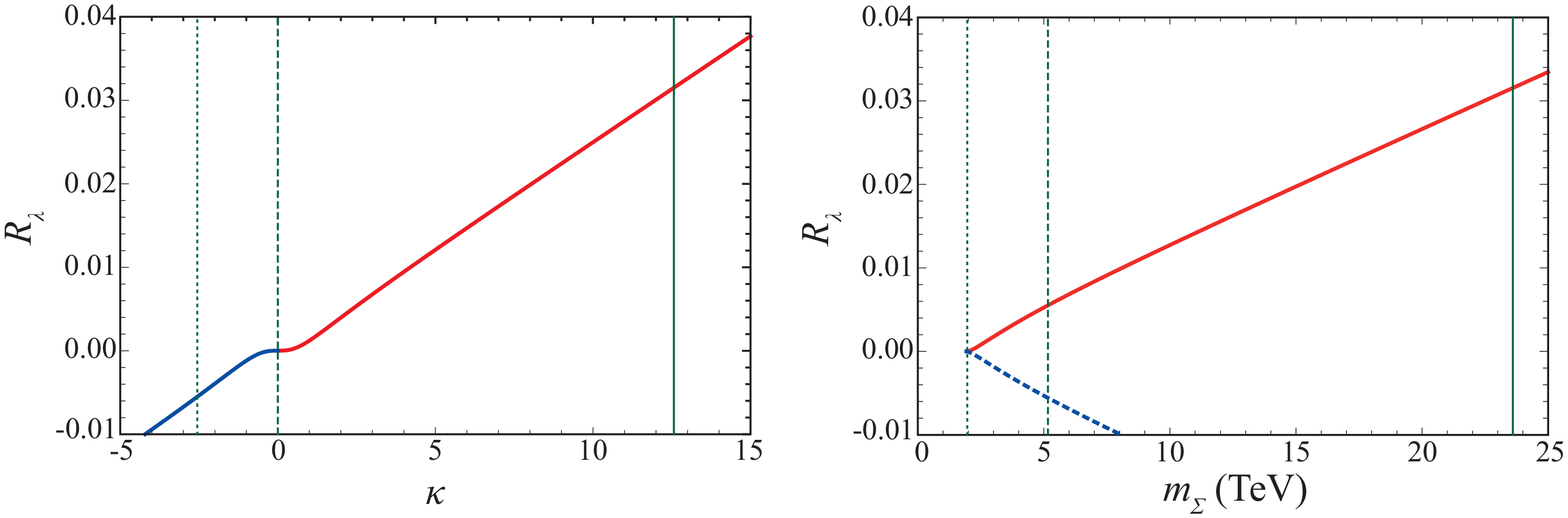}
  \caption{The deviation $R_{\lambda}^{}$ of the trilinear coupling of the SM Higgs boson from its SM value versus the Higgs portal coupling $\kappa$ and the DM mass $m_\Sigma^{}$. In the left panel, the dot, dash and solid vertical lines correspond to $\kappa=-2.6$, $\kappa=0$ and $\kappa=4\pi$, respectively. In the right panel, the solid curve is for $\kappa>0$ and the dot curve is for $\kappa<0$, while the dot, dash and solid vertical lines correspond to $m_\Sigma^{}=2\,\textrm{TeV}$, $m_\Sigma^{}=5.2\,\textrm{TeV}$ and $m_\Sigma^{}=23.6\,\textrm{TeV}$, respectively. We have $R_{\lambda}^{}=-0.014$ for $\kappa=-2.6$ and $m_\Sigma^{}=5.2\,\textrm{TeV}$, $R_{\lambda}^{}=0$ for $\kappa=0$ and $m_\Sigma^{}=2\,\textrm{TeV}$, while $R_{\lambda}^{}=0.032$ for $\kappa=4\pi$ and $m_\Sigma^{}=23.6\,\textrm{TeV}$.}
  \label{h-trilinear}
\end{figure*}

With the presence of the Higgs portal interaction, we can realize a one-loop diagram mediated by the inert Higgs triplet $\Sigma$ to give a dimension-6 operator of the SM Higgs doublet $\phi$ \cite{zhang1993}. By integrating out the inert Higgs triplet $\Sigma$, we obtain
\begin{eqnarray}
V=\mu_\phi^2\phi^\dagger_{}\phi+\lambda(\phi^\dagger_{}\phi)^2_{}+\frac{3\kappa^3_{}}{16\pi^2_{}m_\Sigma^2}(\phi^\dagger_{}\phi)^3_{}\,.
\end{eqnarray}
By minimizing this potential, we have
\begin{eqnarray}
\mu_\phi^2+3\lambda v^2_{}+\frac{9\kappa^3_{}v^4_{}}{64\pi^2_{}m_\Sigma^2}=0\,.
\end{eqnarray}
Then the quadratic and trilinear terms of the Higgs boson $h$ could be extracted,
\begin{eqnarray}
\mathcal{L}&\supset& -\frac{1}{2}m_h^2 h^2_{}-\lambda_{\textrm{eff}}^{} v h^3_{}~~\textrm{with}\nonumber\\
&&m_h^2=2\lambda v^2_{} +\frac{9\kappa^3_{}v^4_{}}{16\pi^2_{}m_\Sigma^2}\,,\nonumber\\
&&\lambda_{\textrm{eff}}^{}=\lambda+\frac{15\kappa^3_{}v^2_{}}{32\pi^2_{}m_\Sigma^2}
=\frac{m_h^2}{2v^2_{}}+\frac{3\kappa^3_{}v^2_{}}{16\pi^2_{}m_\Sigma^2}\,.
\end{eqnarray}
The trilinear coupling of the Higgs boson yields a deviation from the SM value,
\begin{eqnarray}
R_{\lambda}^{}=\frac{\lambda_{\textrm{eff}}^{}-\lambda_{\textrm{SM}}^{}}{\lambda_{\textrm{SM}}^{}}
=\frac{3\kappa^3_{}v^4_{}}{8\pi^2_{}m_\Sigma^2 m_h^2}~~\textrm{with}~~\lambda_{\textrm{SM}}^{}=\frac{m_h^2}{2v^2_{}}\,.
\end{eqnarray}
In Fig. \ref{h-trilinear}, we show the dependence of this deviation $R_{\lambda}^{}$ on the Higgs portal coupling $\kappa$ and the DM mass $m_\Sigma^{}$. We find these deviations are consistent with the experimental limits \cite{mccullough2014}. Specifically, we note $R_{\lambda}^{}=-1.4\%$ for $\kappa=-2.6$ and $m_\Sigma^{}=5.2\,\textrm{TeV}$, $R_{\lambda}^{}=0$ for $\kappa=0$ and $m_\Sigma^{}=2\,\textrm{TeV}$, while $R_{\lambda}^{}=3.2\%$ for $\kappa=4\pi$ and $m_\Sigma^{}=23.6\,\textrm{TeV}$. It is clear now that these numerical results should be in the convincing magnitude to explain why Eq. (\ref{lambda}) is a good approximation to determine the low limit $\kappa=-2.6$.

\begin{figure*}[htp]
 \vspace{0cm} \centering
  \includegraphics[width=15cm]{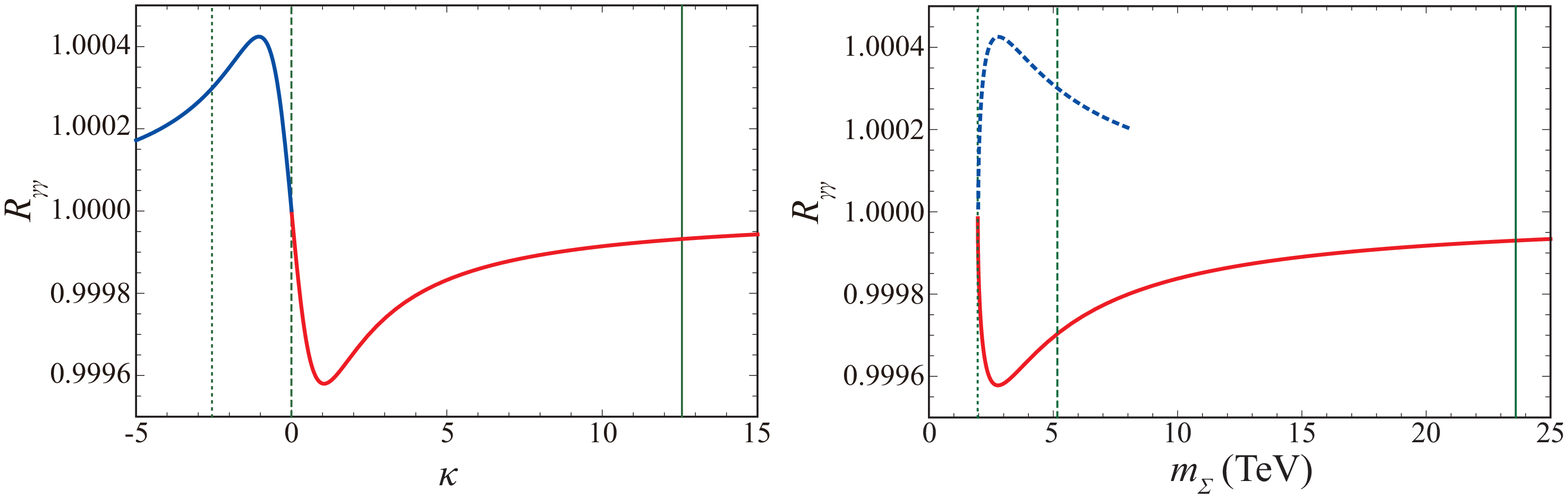}
  \caption{The deviation $R_{\gamma\gamma}^{}$ of the Higgs decay to diphoton from its SM value versus the Higgs portal coupling $\kappa$ and the DM mass $m_\Sigma^{}$. In the left panel, the dot, dash and solid vertical lines correspond to $\kappa=-2.6$, $\kappa=0$ and $\kappa=4\pi$, respectively. In the right panel, the dot curve is for $\kappa<0$ and the solid curve is for $\kappa>0$, while the dot, dash and solid vertical lines correspond to $m_\Sigma^{}=2\,\textrm{TeV}$, $m_\Sigma^{}=5.2\,\textrm{TeV}$ and $m_\Sigma^{}=23.6\,\textrm{TeV}$, respectively. }
  \label{diphoton}
\end{figure*}

We also check the Higgs to diphoton decay \cite{clw2012},
\begin{eqnarray}
R_{\gamma\gamma}^{}&\equiv& \frac{\Gamma\left(h\rightarrow \gamma\gamma\right)}{\Gamma_{\textrm{SM}}\left(h\rightarrow \gamma\gamma\right)} \nonumber \\
&=&\left|1+\frac{\kappa}{2}\frac{v_{}^2}{m_\Sigma^2}\frac{A_0^{}\left(\tau_\Sigma^{}\right)}{A_1^{}\left(\tau_W^{} \right)+\frac{4}{3} A_{\frac{1}{2}}^{}\left(\tau_t^{}\right)}\right|^2~~\textrm{with}\nonumber\\
\tau_X^{}&=&4\frac{m_X^2}{m_h^2}\nonumber\,,\\
A_{0}^{}(x)&=&-x^2_{}\left(\frac{1}{x}-\arcsin^2_{}\frac{1}{\sqrt{x}}\right)\,,\nonumber\\
A_{1}^{}(x)&=&-x^2_{}\left[\frac{2}{x^2_{}}+\frac{3}{x}+3\left(\frac{2}{x}-1\right)\arcsin^2_{}\frac{1}{\sqrt{x}}\right]\,,\nonumber\\
A_{\frac{1}{2}}^{}(x)&=&2x^2_{}\left[\frac{1}{x}+\left(\frac{1}{x}-1\right)\arcsin^2_{}\frac{1}{\sqrt{x}}\right]\,.
\end{eqnarray}
As shown in Fig. \ref{diphoton}, the deviation from the SM prediction is always negligible.

\section{Radiative neutrino masses}

\begin{figure*}
\vspace{6cm} \epsfig{file=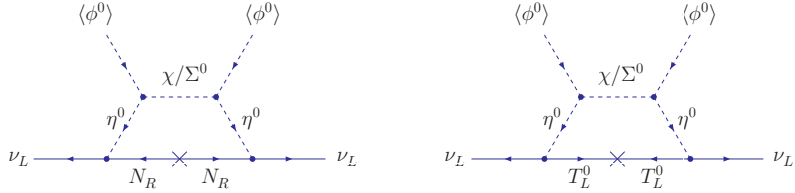, bbllx=6.25cm, bblly=6.0cm,
bburx=16.25cm, bbury=16cm, width=5.5cm, height=5.5cm, angle=0,
clip=0} \vspace{-6.5cm} \caption{\label{numass} The one-loop diagrams for generating the Majorana neutrino masses. }
\end{figure*}

As shown in Fig. \ref{numass}, the left-handed neutrinos can obtain a Majorana mass term,
\begin{eqnarray}
\mathcal{L}\supset -\frac{1}{2}\bar{\nu}_L^{}m_\nu^{}\nu_L^c+\textrm{H.c.}\,,
\end{eqnarray}
after the electroweak symmetry breaking. We take a unitary rotation as below,
\begin{eqnarray}
\left[\begin{array}{c}
\eta_R^0\\
[2.5mm]
\eta_I^0\\
[2.5mm]S
\end{array}\right]=\left[\begin{array}{lll}
U_{\eta^0_R\hat{\eta}^0_R}^{}&U_{\eta^0_R\hat{\eta}^0_I}^{}&U_{\eta^0_R\hat{S}}^{}\\
[2mm]
U_{\eta^0_I\hat{\eta}^0_R}^{}&U_{\eta^0_I\hat{\eta}^0_I}^{}&U_{\eta^0_I\hat{S}}^{}\\
[2mm]
U_{S\hat{\eta}^0_R}^{}&U_{S\hat{\eta}^0_I}^{}&U_{S\hat{S}}^{}
\end{array}\right]\left[\begin{array}{c}
\hat{\eta}_R^0\\
[2.5mm]
\hat{\eta}_I^0\\
[2.5mm]\hat{S}
\end{array}\right]\,,
\end{eqnarray}
to diagonalize the mass matrix of the neutral scalars $\chi/\Sigma$ and $\eta^0_{R,I}$. The neutrino masses then can be exactly computed by
\begin{eqnarray}
\label{numassform}
m_\nu^{}&=&\frac{1}{16\pi^2_{}}yM_{F}^{}\left\{\left[\frac{U_{\eta^0_R\hat{\eta}^0_R}^{}m_{\hat{\eta}^0_{R}}^2
U_{\eta^0_R\hat{\eta}^0_R}^{T}}{m_{\hat{\eta}^0_{R}}^2-M_{F}^2}\ln\left(\frac{m_{\hat{\eta}^0_{R}}^2}{M_{F}^2}\right)\right.\right.\nonumber\\
&& +\frac{U_{\eta^0_R\hat{\eta}^0_I}^{}m_{\hat{\eta}^0_{I}}^2
U_{\eta^0_R\hat{\eta}^0_I}^{T}}{m_{\hat{\eta}^0_{I}}^2-M_{F}^2}\ln\left(\frac{m_{\hat{\eta}^0_{I}}^2}{M_{F}^2}\right)\nonumber\\
&& \left.+\frac{U_{\eta^0_R\hat{S}}^{}m_{\hat{S}}^2
U_{\eta^0_R\hat{S}}^{T}}{m_{\hat{S}}^2-M_{F}^2}\ln\left(\frac{m_{\hat{S}}^2}{M_{F}^2}\right)\right]\nonumber\\
&&-\left[\frac{U_{\eta^0_I\hat{\eta}^0_R}^{}m_{\hat{\eta}^0_{R}}^2
U_{\eta^0_I\hat{\eta}^0_R}^{T}}{m_{\hat{\eta}^0_{R}}^2-M_{F}^2}\ln\left(\frac{m_{\hat{\eta}^0_{R}}^2}{M_{F}^2}\right)\right.\nonumber\\
&& +\frac{U_{\eta^0_I\hat{\eta}^0_I}^{}m_{\hat{\eta}^0_{I}}^2
U_{\eta^0_I\hat{\eta}^0_I}^{T}}{m_{\hat{\eta}^0_{I}}^2-M_{F}^2}\ln\left(\frac{m_{\hat{\eta}^0_{I}}^2}{M_{F}^2}\right)\nonumber\\
&&\left.\left.+\frac{U_{\eta^0_I\hat{S}}^{}m_{\hat{S}}^2
U_{\eta^0_I\hat{S}}^{T}}{m_{\hat{S}}^2-M_{F}^2}\ln\left(\frac{m_{\hat{S}}^2}{M_{F}^2}\right)\right]\right\}y^T_{}\,,
\end{eqnarray}
where $S$ denotes $\chi/\Sigma$ while $F$ standards for $N/T$.

The above formula can be simplified under some limiting conditions. For example, we can obtain
\begin{eqnarray}
(m_\nu^{})_{\alpha\beta}^{}&\simeq&-\frac{v^2_{}}{32\pi^2_{}}\sum_{ijkl}^{}\frac{y_{\alpha i j} y_{\alpha i k} M_{F_i^{}}^{}\rho_{jl}^{}\rho_{kl} ^{}}{M_{\eta_j^{}}^2-M_{\eta_k^{}}^2}\nonumber\\
&&\times \left[\frac{1}{M_{\eta_j^{}}^2-M_{F_i^{}}^2}\ln\left(\frac{M_{F_i^{}}^2}{M_{\eta_j^{}}^2}\right)\right.\nonumber\\
&&\left.
-\frac{1}{M_{\eta_k^{}}^2-M_{F_i^{}}^2}\ln\left(\frac{M_{F_i^{}}^2}{M_{\eta_k^{}}^2}\right)\right]\nonumber\\
[2mm]
&& \textrm{for}~~M_\eta^2\,,~M_F^2\gg M_S^2\,,
\end{eqnarray}
and
\begin{eqnarray}
(m_\nu^{})_{\alpha\beta}^{}&\simeq&-\frac{v^2_{}}{32\pi^2_{}}\sum_{ijkl}^{}\frac{y_{\alpha i j} y_{\alpha i k} M_{F_i^{}}^{}\rho_{jl}^{}\rho_{kl} ^{}}{M_{\eta_j^{}}^2-M_{\eta_k^{}}^2}\nonumber\\
&&\times \left[\frac{1}{M_{\eta_j^{}}^2-M_{S_l^{}}^2}\ln\left(\frac{M_{S_l^{}}^2}{M_{\eta_j^{}}^2}\right)\right.\nonumber\\
&&\left.
-\frac{1}{M_{\eta_k^{}}^2-M_{S_l^{}}^2}\ln\left(\frac{M_{S_l^{}}^2}{M_{\eta_k^{}}^2}\right)\right]\nonumber\\
[2mm]
&& \textrm{for}~~M_\eta^2\,,~M_S^2\gg M_F^2\,.
\end{eqnarray}

By further assuming
\begin{eqnarray}
M_\eta^2\gg M_F^2\,,~M_S^2\,,
\end{eqnarray}
the neutrino masses can have a more simplified form,
\begin{eqnarray}
\label{simplifiednumass}
(m_\nu^{})_{\alpha\beta}^{}&\simeq&-\frac{v^2_{}}{16\pi^2_{}}\sum_{ijkl}^{}\frac{y_{\alpha i j} y_{\alpha i k} M_{F_i^{}}^{}\rho_{jl}^{}\rho_{kl} ^{}}{M_{\eta_j^{}}^2M_{\eta_k^{}}^2}~~\textrm{for}\nonumber\\
&&\ln\left(\frac{M_{\eta}^{}}{M_{F,S}^{}}\right)=\mathcal{O}(1)\,.
\end{eqnarray}
We then can parametrize the Yukawa couplings $y$ by
\begin{eqnarray}
\label{parametrization}
y&=&i\frac{4\pi}{v} U \sqrt{\hat{m}_\nu^{}}O\frac{1}{\rho} M_\eta^2\frac{1}{\sqrt{M_F^{}}} ~~\textrm{with}\nonumber\\
[2mm]
&& m_\nu^{}=U \hat{m}_\nu^{} U^T_{}=U\textrm{diag}\{m_1^{}\,,~m_2^{}\,,~m_3^{}\}U^T_{}\,, \nonumber\\
[2mm]
&&OO^T_{}=O^T_{}O=1\,.
\end{eqnarray}
As the inert Higgs doublets $\eta$ mediate a quartic coupling between the inert Higgs singlets/triplets $\chi/\Sigma$ and the SM Higgs doublet $\phi$, i.e.
\begin{eqnarray}
L\supset -\rho^\dagger_{}\frac{1}{M_\eta^2}\rho \phi^\dagger_{}S^2_{}\phi\,,
\end{eqnarray}
the cubic coupling $\rho$ should favor a perturbative requirement,
\begin{eqnarray}
\rho<\sqrt{4\pi}M_{\eta}^{}\,.
\end{eqnarray}
Applying this constraint to Eq. (\ref{simplifiednumass}), we find
\begin{eqnarray}
M_\eta^2&\lesssim &\frac{v^2_{}}{m_\nu^{}}M_F^{}=(2.5\times 10^{13}_{}\,\textrm{GeV})^2_{}\left(\frac{M_F^{}/ 10^{12}_{}\,\textrm{GeV}}{m_\nu^{}/0.1\,\textrm{eV}}\right)\nonumber\\
&&\textrm{for}~~y<\sqrt{4\pi}\,.
\end{eqnarray}

Note the simple formula (\ref{simplifiednumass}) means the models should contain at least (i) one inert fermion singlet/triplet, one inert Higgs singlet/triplet and two inert Higgs doublets; (ii) one inert fermion singlet/triplet, two inert Higgs singlets/triplets and one inert Higgs doublet; (iii) two inert fermion singlets/triplets, one inert Higgs singlet/triplet and one inert Higgs doublet, in order to give two or more non-zero neutrino mass eigenvalues.

\section{Leptogenesis}

It is well known that we can realize a leptogenesis through the CP-violating decays of the inert fermion singlets/triplets into the inert Higgs doublets and the SM lepton doublets if these inert fermions are heavy enough \cite{ma2006}. Another possibility also holds for the decays of the inert Higgs singlets/triplets into the inert Higgs doublets and the SM Higgs doublet. The produced asymmetry stored in the inert Higgs doublet pairs will eventually turn into a lepton asymmetry stored in the SM leptons after the inert Higgs doublets decay into the SM lepton doublets and the inert fermion singlets/triplets. Alternatively, we can implement a leptogenesis making use of the decays of the inert Higgs doublets. We will focus on this leptogenesis scenario and illustrate its main aspects in the following.

\begin{figure*}
\vspace{7.5cm} \epsfig{file=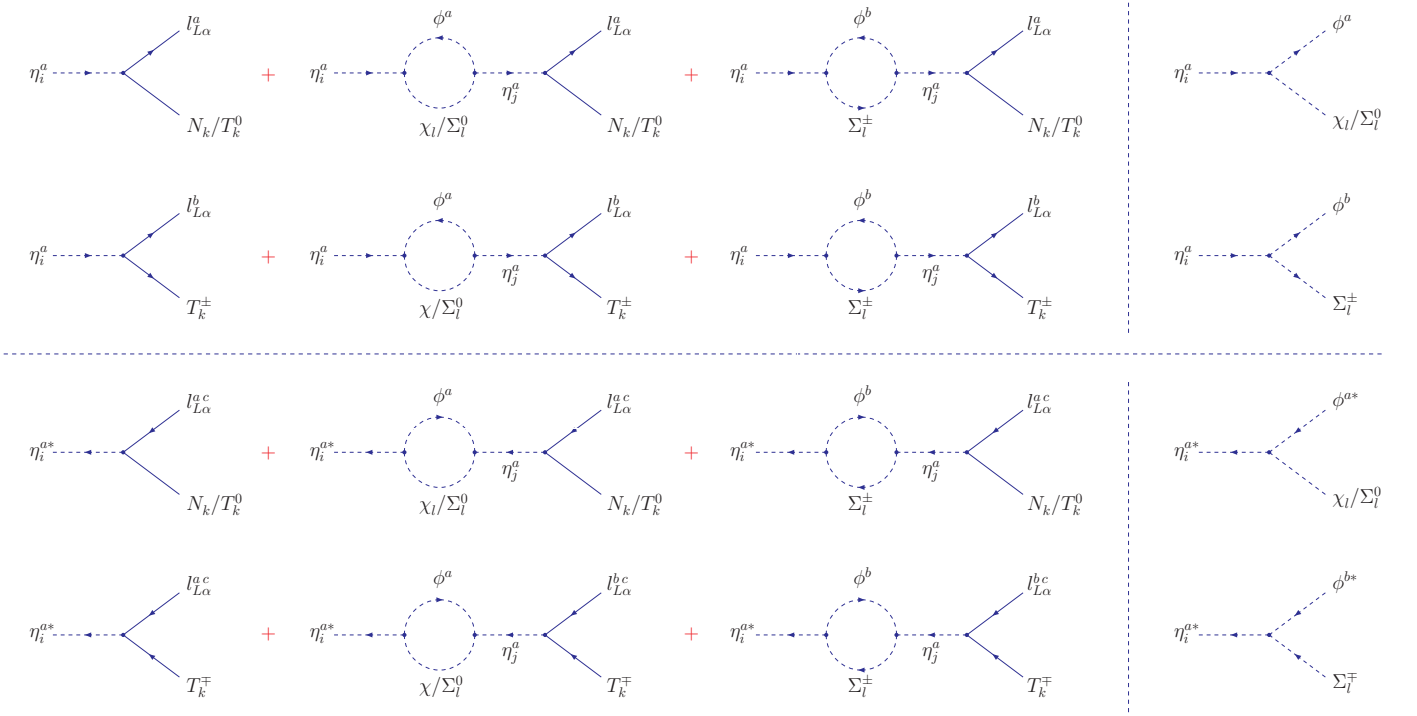, bbllx=15.6cm, bblly=6.0cm,
bburx=25.6cm, bbury=16cm, width=5.3cm, height=5.3cm, angle=0,
clip=0} \vspace{-2.7cm} \caption{\label{HDdecay} The decays of the inert Higgs doublets $\eta$ into the inert Higgs singlets/triplets $\chi/\Sigma$, the inert fermion singlets/triplets $N/T$ as well as the SM Higgs doublet $\phi$ and the SM lepton doublets $l_L^{}$.}
\end{figure*}

As shown in Fig. \ref{HDdecay}, the inert Higgs doublets can have two decay channels: one is into the inert Higgs singlets/triplets and the SM Higgs doublet, the other is into the SM lepton doublets and the inert fermion singlets/triplets. Therefore, the decays of the inert Higgs doublets can generate a lepton asymmetry stored in the SM leptons.

We calculate the decay widths at tree level and the CP asymmetries at one-loop order in the SSD, STD, TSD and TTD models, respectively.
\begin{itemize}
\item In the SSD model,
\begin{eqnarray}
\Gamma_{\eta_i^{}}^{}&\equiv&\Gamma_{\eta_i^{\ast}}^{}=
\frac{1}{16\pi}M_{\eta_i^{}}^{}\left[(y^\dagger_{}y)_{ii}^{}+\frac{(\rho\rho^\dagger_{})_{ii}^{}}{M_{\eta_i^{}}^2}\right]\nonumber\\
&&\quad \quad \!\geq \frac{1}{8\pi}\sqrt{(y^\dagger_{}y)_{ii}^{}(\rho\rho^\dagger_{})_{ii}^{}}~~\textrm{with}\nonumber\\
[2mm]
\Gamma_{\eta_i^{}}^{}&\equiv&\sum_{\alpha j}[\Gamma(\eta_i^{}\rightarrow l_{L\alpha}^{}+N_{Rj}^{c})+\Gamma(\eta_i^{}\rightarrow \phi+\chi_j^{})]\,,\nonumber\\
\Gamma_{\eta_i^{\ast}}^{}&\equiv&\sum_{\alpha j}[\Gamma(\eta_i^{\ast}\rightarrow l_{L\alpha}^{c}+N_{Rj}^{})+\Gamma(\eta_i^{\ast}\rightarrow \phi_{}^{\ast}+\chi_j^{})]\,,\nonumber\\
&&\end{eqnarray}
\begin{eqnarray}
\varepsilon_{\eta_i^{}}^{}&\equiv&\frac{\sum_{\alpha j}[\Gamma(\eta_i^{}\rightarrow l_{L\alpha}^{}+N_{Rj}^{c})-\Gamma(\eta_i^{}\rightarrow l_{L\alpha}^{c}+N_{Rj}^{})]}{\Gamma_{\eta_i^{}}^{}}\nonumber\\
&=&\frac{1}{4\pi}\sum_{j\neq i}^{}\frac{\textrm{Im}[(y^\dagger_{}y)_{ij}^{}(\rho\rho^\dagger_{})_{ji}^{}]}
{(y^\dagger_{}y)_{ii}^{}+\frac{(\rho\rho^\dagger_{})_{ii}^{}}{M_{\eta_i^{}}^2}}\frac{1}{M_{\eta_i^{}}^{2}-M_{\eta_j^{}}^{2}}\nonumber\\
&\leq&\frac{1}{8\pi}\sum_{j\neq i}^{}\frac{\textrm{Im}[(y^\dagger_{}y)_{ij}^{}(\rho\rho^\dagger_{})_{ji}^{}]}
{\sqrt{(y^\dagger_{}y)_{ii}^{}(\rho\rho^\dagger_{})_{ii}^{}}}\frac{M_{\eta_i^{}}^{}}{M_{\eta_i^{}}^{2}-M_{\eta_j^{}}^{2}}
\end{eqnarray}
\item In the STD model,
\begin{eqnarray}
\Gamma_{\eta_i^{}}^{}&\equiv&\Gamma_{\eta_i^{\ast}}^{}=
\frac{1}{16\pi}M_{\eta_i^{}}^{}\left[(y^\dagger_{}y)_{ii}^{}+3\frac{(\rho\rho^\dagger_{})_{ii}^{}}{M_{\eta_i^{}}^2}\right]\nonumber\\
&&\quad \quad \!\geq \frac{\sqrt{3}}{8\pi}\sqrt{(y^\dagger_{}y)_{ii}^{}(\rho\rho^\dagger_{})_{ii}^{}}~~\textrm{with}\nonumber\\
[2mm]
\Gamma_{\eta_i^{}}^{}&\equiv&\sum_{\alpha j}[\Gamma(\eta_i^{}\rightarrow l_{L\alpha}^{}+N_{Rj}^{c})+\Gamma(\eta_i^{}\rightarrow \phi+\Sigma_j^{})]\,,\nonumber\\
\Gamma_{\eta_i^{\ast}}^{}&\equiv&\sum_{\alpha j}[\Gamma(\eta_i^{\ast}\rightarrow l_{L\alpha}^{c}+N_{Rj}^{})+\Gamma(\eta_i^{\ast}\rightarrow \phi_{}^{\ast}+\Sigma_j^{})]\,,\nonumber\\
&&\end{eqnarray}
\begin{eqnarray}
\varepsilon_{\eta_i^{}}^{}&\equiv&\frac{\sum_{\alpha j}[\Gamma(\eta_i^{}\rightarrow l_{L\alpha}^{}+N_{Rj}^{c})-\Gamma(\eta_i^{}\rightarrow l_{L\alpha}^{c}+N_{Rj}^{})]}{\Gamma_{\eta_i^{}}^{}}\nonumber\\
&=&\frac{3}{4\pi}\sum_{j\neq i}^{}\frac{\textrm{Im}[(y^\dagger_{}y)_{ij}^{}(\rho\rho^\dagger_{})_{ji}^{}]}
{(y^\dagger_{}y)_{ii}^{}+3\frac{(\rho\rho^\dagger_{})_{ii}^{}}{M_{\eta_i^{}}^2}}\frac{1}{M_{\eta_i^{}}^{2}-M_{\eta_j^{}}^{2}}\nonumber\\
&\leq&\frac{\sqrt{3}}{8\pi}\sum_{j\neq i}^{}\frac{\textrm{Im}[(y^\dagger_{}y)_{ij}^{}(\rho\rho^\dagger_{})_{ji}^{}]}
{\sqrt{(y^\dagger_{}y)_{ii}^{}(\rho\rho^\dagger_{})_{ii}^{}}}\frac{M_{\eta_i^{}}^{}}{M_{\eta_i^{}}^{2}-M_{\eta_j^{}}^{2}}\,.
\end{eqnarray}
\item In the TSD model,
\begin{eqnarray}
\Gamma_{\eta_i^{}}^{}&\equiv&\Gamma_{\eta_i^{\ast}}^{}=
\frac{1}{16\pi}M_{\eta_i^{}}^{}\left[3(y^\dagger_{}y)_{ii}^{}+\frac{(\rho\rho^\dagger_{})_{ii}^{}}{M_{\eta_i^{}}^2}\right]\nonumber\\
&&\quad \quad \!\geq \frac{\sqrt{3}}{8\pi}\sqrt{(y^\dagger_{}y)_{ii}^{}(\rho\rho^\dagger_{})_{ii}^{}}~~\textrm{with}\nonumber\\
[2mm]
\Gamma_{\eta_i^{}}^{}&\equiv&\sum_{\alpha j}[\Gamma(\eta_i^{}\rightarrow l_{L\alpha}^{}+T_{Lj}^{})+\Gamma(\eta_i^{}\rightarrow \phi+\chi_j^{})]\,,\nonumber\\
\Gamma_{\eta_i^{\ast}}^{}&\equiv&\sum_{\alpha j}[\Gamma(\eta_i^{\ast}\rightarrow l_{L\alpha}^{c}+T_{Lj}^{c})+\Gamma(\eta_i^{\ast}\rightarrow \phi_{}^{\ast}+\chi_j^{})]\,,\nonumber\\
&&\end{eqnarray}
\begin{eqnarray}
\varepsilon_{\eta_i^{}}^{}&\equiv&\frac{\sum_{\alpha j}[\Gamma(\eta_i^{}\rightarrow l_{L\alpha}^{}+T_{Lj}^{})-\Gamma(\eta_i^{}\rightarrow l_{L\alpha}^{c}+T_{Lj}^{c})]}{\Gamma_{\eta_i^{}}^{}}\nonumber\\
&=&\frac{3}{4\pi}\sum_{j\neq i}^{}\frac{\textrm{Im}[(y^\dagger_{}y)_{ij}^{}(\rho\rho^\dagger_{})_{ji}^{}]}
{3(y^\dagger_{}y)_{ii}^{}+\frac{(\rho\rho^\dagger_{})_{ii}^{}}{M_{\eta_i^{}}^2}}\frac{1}{M_{\eta_i^{}}^{2}-M_{\eta_j^{}}^{2}}\nonumber\\
&\leq&\frac{\sqrt{3}}{8\pi}\sum_{j\neq i}^{}\frac{\textrm{Im}[(y^\dagger_{}y)_{ij}^{}(\rho\rho^\dagger_{})_{ji}^{}]}
{\sqrt{(y^\dagger_{}y)_{ii}^{}(\rho\rho^\dagger_{})_{ii}^{}}}\frac{M_{\eta_i^{}}^{}}{M_{\eta_i^{}}^{2}-M_{\eta_j^{}}^{2}}\,.
\end{eqnarray}
\item In the TTD model,
\begin{eqnarray}
\Gamma_{\eta_i^{}}^{}&\equiv&\Gamma_{\eta_i^{\ast}}^{}=
\frac{3}{16\pi}M_{\eta_i^{}}^{}\left[(y^\dagger_{}y)_{ii}^{}+\frac{(\rho\rho^\dagger_{})_{ii}^{}}{M_{\eta_i^{}}^2}\right]\nonumber\\
&&\quad \quad \!\geq \frac{3}{8\pi}\sqrt{(y^\dagger_{}y)_{ii}^{}(\rho\rho^\dagger_{})_{ii}^{}}~~\textrm{with}\nonumber\\
[2mm]
\Gamma_{\eta_i^{}}^{}&\equiv&\sum_{\alpha j}[\Gamma(\eta_i^{}\rightarrow l_{L\alpha}^{}+T_{Lj}^{})+\Gamma(\eta_i^{}\rightarrow \phi+\Sigma_j^{})]\,,\nonumber\\
\Gamma_{\eta_i^{\ast}}^{}&\equiv&\sum_{\alpha j}[\Gamma(\eta_i^{\ast}\rightarrow l_{L\alpha}^{c}+T_{Lj}^{c})+\Gamma(\eta_i^{\ast}\rightarrow \phi_{}^{\ast}+\Sigma_j^{})]\,,\nonumber\\
&&\end{eqnarray}
\begin{eqnarray}
\varepsilon_{\eta_i^{}}^{}&\equiv&\frac{\sum_{\alpha j}[\Gamma(\eta_i^{}\rightarrow l_{L\alpha}^{}+T_{Lj}^{})-\Gamma(\eta_i^{}\rightarrow l_{L\alpha}^{c}+T_{Lj}^{c})]}{\Gamma_{\eta_i^{}}^{}}\nonumber\\
&=&\frac{3}{4\pi}\sum_{j\neq i}^{}\frac{\textrm{Im}[(y^\dagger_{}y)_{ij}^{}(\rho\rho^\dagger_{})_{ji}^{}]}
{(y^\dagger_{}y)_{ii}^{}+\frac{(\rho\rho^\dagger_{})_{ii}^{}}{M_{\eta_i^{}}^2}}\frac{1}{M_{\eta_i^{}}^{2}-M_{\eta_j^{}}^{2}}\nonumber\\
&\leq&\frac{3}{8\pi}\sum_{j\neq i}^{}\frac{\textrm{Im}[(y^\dagger_{}y)_{ij}^{}(\rho\rho^\dagger_{})_{ji}^{}]}
{\sqrt{(y^\dagger_{}y)_{ii}^{}(\rho\rho^\dagger_{})_{ii}^{}}}\frac{M_{\eta_i^{}}^{}}{M_{\eta_i^{}}^{2}-M_{\eta_j^{}}^{2}}\,.
\end{eqnarray}
\end{itemize}

By making use of the parametrization (\ref{parametrization}), the above decay widths and CP asymmetries may be simplified. For example, in the models with one inert fermion singlet/triplet, one inert Higgs singlet/triplet and two or more inert Higgs doublets, we can derive
\begin{eqnarray}
\label{width}
\Gamma_{\eta_i^{}}^{}&\geq&\frac{C}{2} \frac{M_{\eta_i^{}}^2}{v}\sqrt{\frac{(O^\dagger_{}\hat{m}_\nu^{}O)_{ii}^{}}{M_F^{}}}=
\frac{C}{2} \frac{M_{\eta_i^{}}^2}{v}\sqrt{\frac{\tilde{m}_i^{}}{M_F^{}}}\,,~~\\
\label{cp}
|\varepsilon_{\eta_i^{}}^{}|&<&\varepsilon_{\eta_i^{}}^{\textrm{max}}= \frac{C}{2}\frac{M_{\eta_i^{}}^{}}{v}\sqrt{\frac{m_{\textrm{max}}^{}}{M_F^{}}}
~~\textrm{for}~~M_{\eta_i^{}}^{2}\ll M_{\eta_j^{}}^{2}\,,~~
\end{eqnarray}
where the parameters $C$, $\tilde{m}_i^{}$ and $m_{\textrm{max}}^{}$ are defined by
\begin{eqnarray}
&&m_{\textrm{max}}^{}=\max \{m_1^{}\,,~m_2^{}\,,~m_3^{}\}\,,\nonumber\\
[2mm]
&&m_{\textrm{min}}^{}=\min \{m_1^{}\,,~m_2^{}\,,~m_3^{}\}\,,\nonumber\\
[2mm]
&&\tilde{m}_i^{}\equiv (O^\dagger_{}\hat{m}_\nu^{}O)_{ii}^{}\in (m_{\textrm{min}}^{}\,,~m_{\textrm{max}}^{})\,,\nonumber\\
[2mm]
&&C=\left\{\begin{array}{cl}1 &\textrm{in~the~SSD~models}\,,\\
[2mm]
 \sqrt{3} &\textrm{in~the~STD/TSD~models}\,,\\
 [2mm]
 3 &\textrm{in~the~TTD~models}\,.
\end{array}\right.
\end{eqnarray}

Instead of fully integrating the Boltzmann equations to determine the final baryon asymmetry, we adopt an instructive and reliable estimation for demonstration, assuming a hierarchical spectrum for the inert Higgs doublets. Consequently, the final baryon asymmetry should be mostly produced by the decays of the lightest inert Higgs doublet $\eta_1^{}$. We define
\begin{eqnarray}
\label{rwidth}
K&=&\frac{\Gamma_{\eta_1^{}}^{}}{2H(T)}\left|_{T=M_{\eta_1^{}}^{}}^{}\right.\,,
\end{eqnarray}
where $H(T)$ is the Hubble constant,
\begin{eqnarray}
H=\left(\frac{8\pi^{3}_{}g_{\ast}^{}}{90}\right)^{\frac{1}{2}}_{}
\frac{T^{2}_{}}{M_{\textrm{Pl}}^{}}\,,
\end{eqnarray}
with $g_{\ast}^{}$ being the relativistic degrees of freedom during the leptogenesis epoch. For $1\ll K \lesssim 10^6_{}$, the final baryon asymmetry can well approximate to \cite{kt1990}
\begin{eqnarray}
\eta_B^{}&=&\frac{n_B^{}}{s}\simeq -\frac{28}{79}\times \frac{\varepsilon_{\eta_1^{}}^{}}{g_\ast^{}K z_f^{}} \times 2 \nonumber\\
&&\textrm{with}~~z_f^{}=\frac{M_{\eta_1^{}}^{}}{T_f^{}}\simeq 4.2(\ln K)^{0.6}_{}\,.
\end{eqnarray}
Here $n_B^{}$ and $s$, respectively, are the baryon number density and the entropy density, $T_{f}^{}$ corresponds to the temperature when the processes damping the lepton asymmetry freeze out, the factor $-\frac{28}{79}$ is the sphaleron lepton-to-baryon coefficient, while the factor $2$ appears because the decaying particle $\eta_{1}^{}$ is a doublet.

To provide a numerical illustration, we consider the STD model with one inert fermion singlet, one inert Higgs triplet, two or more inert Higgs doublets. After taking $g_\ast^{}=109.75$ (the SM fields plus one inert Higgs triplet) and setting the inputs,
\begin{eqnarray}
&&M_{\eta_1^{}}^{}=2\times 10^{13}_{}\,\textrm{GeV}\,,~~M_F^{}=2\times 10^{12}_{}\,\textrm{GeV}\,,\nonumber\\
&&m_{\textrm{max}}^{}=0.05\,\textrm{eV}\,,~~\tilde{m}_{11}^{}=0.01\,\textrm{eV}\,,
\end{eqnarray}
in Eqs. (\ref{width}), (\ref{cp}) and (\ref{rwidth}), we read
\begin{eqnarray}
&&\varepsilon_{\eta_i^{}}^{\textrm{max}}=0.352\,,~~K=2761\,,\nonumber\\
&&z_f^{}=14.6\,,~~T_f^{}=1.4\times 10^{12}_{}\,\textrm{GeV}\,,
\end{eqnarray}
and then obtain an expected baryon asymmetry \cite{olive2014},
\begin{eqnarray}
\eta_B^{}=10^{-10}_{}\left(\frac{\varepsilon_{\eta_1^{}}^{}}{-0.002\,\varepsilon_{\eta_1^{}}^{\textrm{max}}}\right)\,.
\end{eqnarray}

\section{Summary}

In this paper, we have built a class of models by introducing the inert fermion singlets/triplets, the inert Higgs singlets/triplets and the inert Higgs doublets. In our models, the Majorana neutrino masses could only be induced through a one-loop diagram mediated by these inert fields, as a consequence of the softly broken lepton number and the exactly conserved $Z_2^{}$ discrete symmetry. The interactions for generating the neutrino masses can also accommodate the decays of the heavier inert fields into the lighter ones and the SM fields. Such decays can realize a successful leptogenesis to explain the baryon asymmetry in the universe. As an example, we have considered the inert Higgs doublet decays. While on the other side, the lightest inert field could provide a stable DM candidate. We have performed a systematic study on the inert Higgs triplet DM scenario with emphasis on investigating some phenomenological effects from the Higgs portal interaction. Our computation shows the interference between the Higgs portal and gauge interactions can result in a drastic decrease of the DM-nucleon scattering cross section.

\textbf{Acknowledgement}: This work was supported by the Recruitment Program for Young Professionals under Grant No. 15Z127060004, the Shanghai Jiao Tong University under Grant No. WF220407201 and the Shanghai Laboratory for Particle Physics and Cosmology under Grant No. 11DZ2260700.

\end{document}